# Patterns of Transposable Element Distribution Around Chromatin Ligation Points Revealed by Micro-C Data Analysis

Alexandr V. Vikhorev[1], Michael M. Rempel[1], Oksana O. Polesskaya[1], Ivan V. Savelev[1] and Max V. Myakishev-Rempel[1]

DNA Resonance Research Foundation, dnaresonance.org,

1. DNA Resonance Research Foundation, San Diego, CA, USA,

Correspondence: MMR - [max@dnaresonance.org](max@dnaresonance.org)

August 19, 2024

## Abstract

Transposable elements (TEs) constitute a significant portion of eukaryotic genomes, yet their role in chromatin organization remains poorly understood. This study investigates the distribution patterns of TEs around chromatin ligation points (LPs) identified through Micro-C experiments in human cells. We analyzed the density of various TE families within a 100kb window centered on LPs, focusing on major families such as Alu and LINE-1 (L1) elements. Our findings reveal distinct, non-random distribution patterns that differ between TE families and exhibit consistent strand-specific biases. These patterns were reproducible across two independent datasets and showed marked differences from random genomic distributions. Notably, we observed family-specific variations in TE density near LPs, with some families showing depletion at LPs followed by periodic fluctuations in density. The consistency of these patterns across TE families and their orientation relative to chromosome arms suggest a fundamental relationship between TEs and higher-order chromatin structure. Our results provide new insights into the potential role of TEs in genome organization and challenge the notion of TEs as passive genomic components. This study lays the groundwork for future investigations into the functional implications of TE distribution in chromatin architecture and gene regulation.

## Introduction

Transposable elements (TEs) constitute a significant portion of eukaryotic genomes, often comprising more than half of the genomic content in many species. Once considered "junk DNA," these mobile genetic elements have been increasingly implicated in various aspects of genome function and evolution. Recent advancements in chromosome conformation capture techniques, particularly Micro-C, have enabled high-resolution mapping of chromatin interactions, providing unprecedented insights into the three-dimensional organization of the genome.

The rationale behind our investigation stems from the hypothesis that repetitive elements, due to their sequence homology, might facilitate or be influenced by chromatin folding. This hypothesis is rooted in the observation that homologous sequences have the potential to form physical interactions in the nuclear space. By examining the distribution of TEs around ligation points (LPs) identified through Micro-C experiments, we sought to uncover patterns that could shed light on the role of TEs in chromatin organization and potentially in the formation of chromatin loops.

Our study focuses on two of the most abundant TE families in the human genome: Alu elements, short interspersed nuclear elements (SINEs) approximately 300 base pairs in length, and LINE-1 (L1) elements, long interspersed nuclear elements that can span several kilobases. These TE families have distinct evolutionary histories and genomic distributions, making them ideal candidates for comparative analysis in the context of chromatin structure.

## Methods

We analyzed Micro-C data from two independent datasets to identify ligation points (LPs). The Micro-C data were obtained from public micro-C data deposited on SRA database from NCBI SRR12625672 and SRR12625674.
https://www.ncbi.nlm.nih.gov/sra/?term=SRR12625672
https://www.ncbi.nlm.nih.gov/sra/?term=SRR12625674
The data corresponds to 42.95M paired-end reads with 150nt each. The data has high quality from HUDEP cell line. In LigP finder v2, it performs a whole-genome alignment.

The python code can be downloaded from
https://github.com/maxrempel/DRRF/tree/main/LigP_finder-main.

The Ligation points, LPs, were obtained with hg38, unmasked version:
https://hgdownload.soe.ucsc.edu/goldenPath/hg38/bigZips/hg38.fa.gz

We used the gold standard hg38 TE annotation publicly available on UCSC database:
https://hgdownload.soe.ucsc.edu/goldenPath/hg38/bigZips/hg38.fa.out.gz

For each identified LP, we examined the genomic regions extending 50kb in both directions, which we termed "exbors" (extended borders). Within these regions, we cataloged the presence and orientation of transposable elements using bedtools, function intersect. We focused on the top 20 most abundant TE families based on their genomic copy numbers.

To quantify TE distribution, we calculated TE density in 2kb bins across the 100kb window centered on each LP. Importantly, we conducted separate analyses for elements on the plus and minus strands to investigate potential strand-specific patterns. To visualize these distribution patterns, we generated Kernel Density Estimation (KDE) plots, which provide a smooth representation of the data while preserving important features.

To ensure the biological significance of our observations, we performed random control analyses. This involved generating random LP positions across the genome and comparing TE distributions around these points to the distributions observed around real LPs. This control step was crucial in distinguishing genuine biological patterns from background genomic noise.

## Results

Our analysis revealed complex patterns of TE distribution around chromatin ligation points, with notable similarities across various TE families and subfamilies. We examined the density of six major TE families: Alu, L1 (LINE-1), L2 (LINE-2), MER, MIR, and MLT elements, as well as several subfamilies within these groups, within a 100 kb window centered on LPs identified through Micro-C experiments. Figure 1 presents the distribution of L1 elements around LPs (ligation points), which is largely representative of the trends seen across all examined TE families.

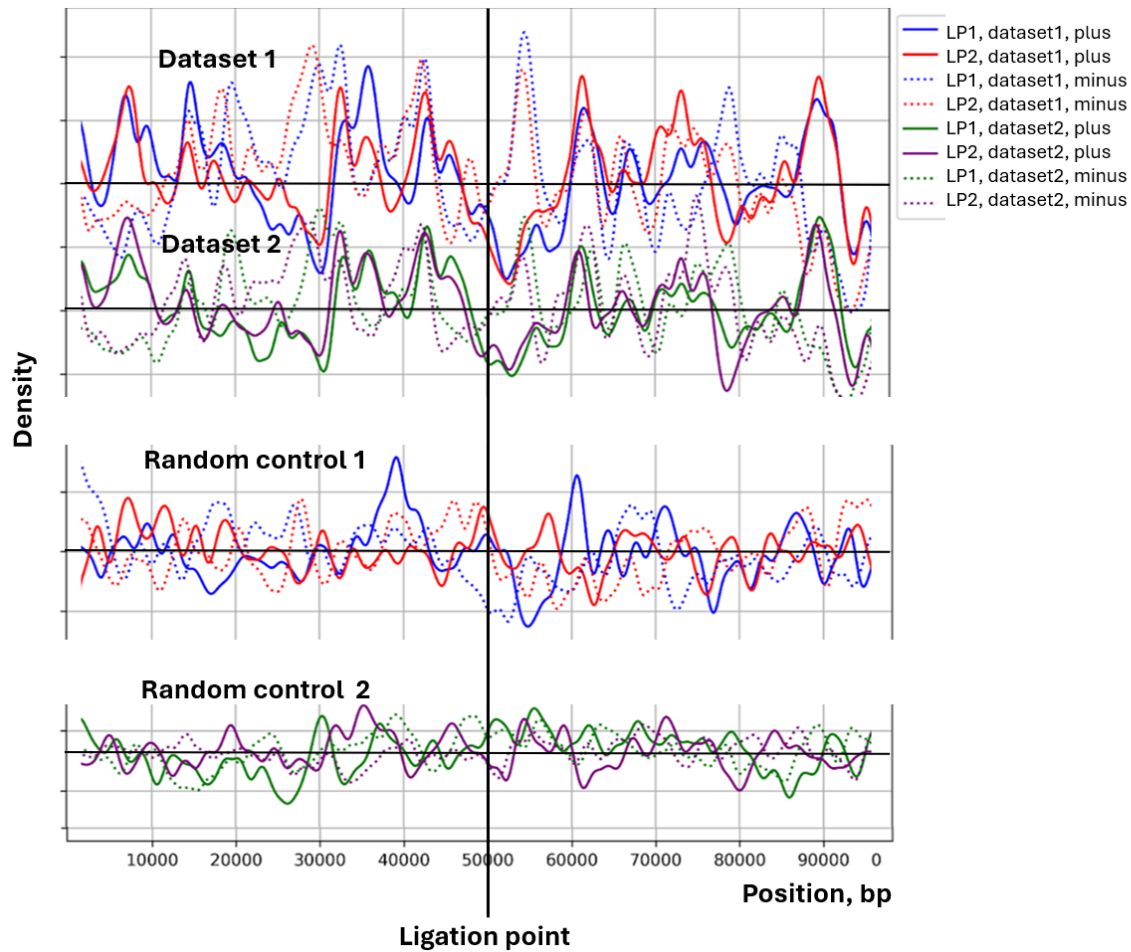

Figure 1: Distribution of L1 transposable elements around chromatin ligation point. Density of L1 elements (a major subfamily of Long Interspersed Nuclear Elements or LINEs) relative to chromatin ligation points (LPs) in two experimental Micro-C datasets and their corresponding random controls. The x-axis represents the position relative to the LP (vertical black line at 50,000 bp), spanning 100 kb. The y-axis shows the density of L1 elements (count of L1 elements divided by the bin size, bp). Solid lines represent L1 elements in plus strand (therefore oriented left to right), dotted lines represent the opposite oriented L1 elements, labeled as minus strand. Dataset 1 (blue/red) and Dataset 2 (green/purple) show experimental data for LP1 and LP2 regions. LP1 and LP2 ligation points were numbered from left (start) to right (end) of the chromosome. Random controls 1 and 2 correspond to Datasets 1 and 2 respectively, generated using randomized LP positions. The graphs are oriented from left to right on the chromosome.

Figure 1 presents a view of L1 transposable element distribution around chromatin ligation points (LPs) identified through Micro-C experiments. L1, or LINE-1, is a major subfamily of Long Interspersed Nuclear Elements (LINEs) comprising about 20% of the human genome. The data reveal striking patterns that suggest a non-random association between L1 elements and chromatin structure.

Key observations from the figure include:

1. Correlation between datasets and LPs: There is a remarkable similarity in the distribution patterns between Dataset 1 and Dataset 2, as well as between LP1 and LP2. This consistency across independent datasets and different ligation points strengthens the biological significance of the observed patterns.
2. Strand bias: In both datasets, there is a clear strand bias, with the plus strand (solid lines) showing generally higher L1 density compared to the minus strand (dotted lines). This asymmetry could reflect strand-specific roles of L1 elements in chromatin organization or biases in their genomic integration and retention.
3. Periodic fluctuations: The L1 density shows pronounced, periodic fluctuations across the 100 kb window. These fluctuations are more prominent in Dataset 1 and appear to be roughly symmetrical around the ligation point.
4. Depletion at the ligation point: There is a noticeable dip in L1 density directly at the ligation point (50,000 bp), particularly evident in Dataset 1. This depletion could indicate exclusion of L1 elements from immediate chromatin interaction sites.
5. Random controls: The random control datasets show markedly different patterns characterized by less pronounced fluctuations, reduced strand bias, and lack of correlation between curves. This contrast underscores the biological significance of the patterns observed in the experimental data.

These findings collectively suggest a complex interplay between L1 elements and higher-order chromatin structure. The observed patterns could result from selective pressures on L1 insertions, preferential retention in specific chromatin environments, or a direct functional role of L1 elements in shaping chromatin architecture.

This figure provides compelling evidence for the non-random distribution of L1 elements in relation to chromatin architecture. It sets the stage for further investigations into the functional implications of these patterns, including potential roles in chromatin loop formation or the establishment of topologically associating domains (TADs). Future studies may focus on understanding the mechanisms driving these distribution patterns and their potential impact on gene regulation and genome organization.

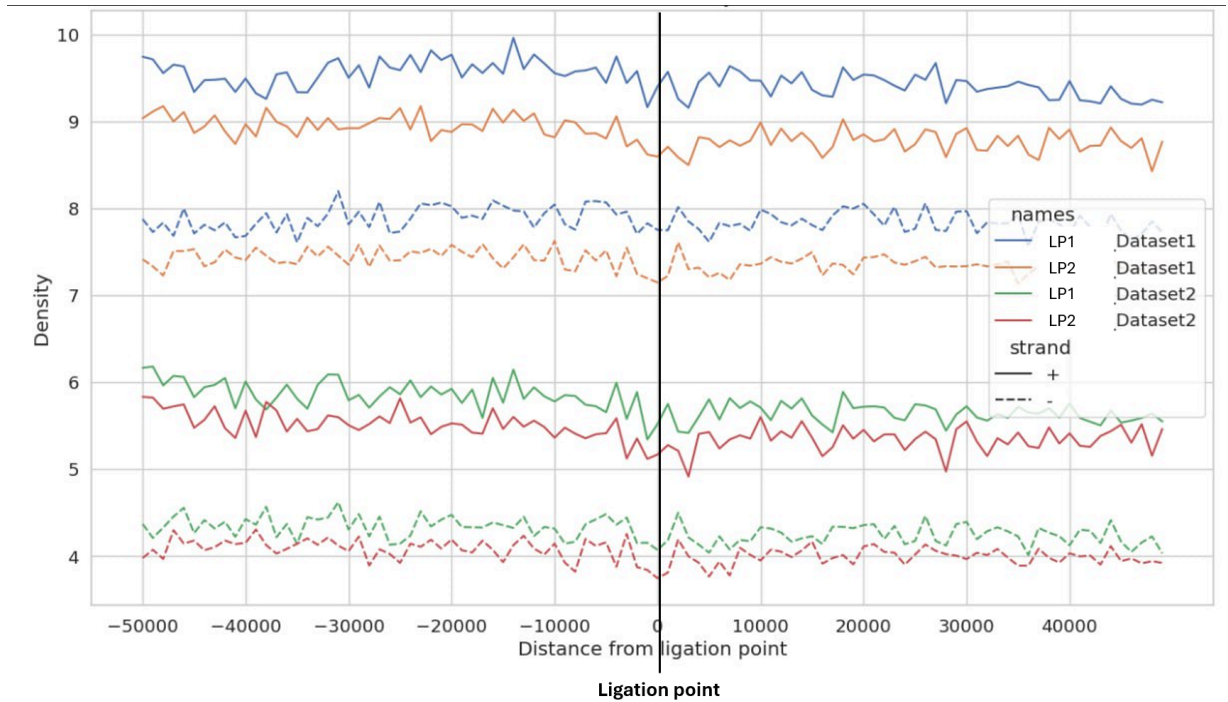

**Figure 2: Distribution of Alu transposable elements around chromatin ligation points.**
Density of Alu elements (a subfamily of Short Interspersed Nuclear Elements or SINEs) relative to chromatin ligation points (LPs) in two experimental datasets. The x-axis represents the distance from the LP (at 0), spanning 100 kb (-50,000 to +50,000 bp). The y-axis shows the density of Alu elements. Solid lines represent plus strand, dotted lines represent minus strand. Dataset 1 (blue/orange) and Dataset 2 (green/red) show data for LP1 and LP2 regions. The y-axis shifts were added artificially to prevent overlap of the curves.

Figure 2 presents the distribution of Alu elements around chromatin ligation points (LPs) identified through Micro-C experiments. Alu elements, a subfamily of Short Interspersed Nuclear Elements (SINEs), comprise approximately 11% of the human genome, making their distribution particularly relevant for understanding genome organization.

Key observations from Figure 2 reveal striking patterns in Alu element distribution around chromatin ligation points. The data show remarkable consistency across datasets and ligation points, with Dataset 1 and Dataset 2 exhibiting highly similar distribution patterns, as do LP1 and LP2. This reproducibility not only reinforces the biological significance of the observations but also aligns with trends observed in other transposable element families. Additionally, a clear strand bias is evident throughout the data, with the plus strand (represented by solid lines) consistently displaying higher Alu density compared to the minus strand (dotted lines). This asymmetry is a consistent feature in both datasets and mirrors the strand bias observed in other TE families, suggesting a fundamental principle in the relationship between TE orientation and chromatin structure. The consistency of these patterns across different experimental conditions and genomic contexts underscores the non-random nature of Alu element distribution in relation to chromatin architecture.

The observed patterns suggest that Alu elements, despite their high frequency, have non-random associations with chromatin architecture. The consistent strand bias across different TE families points to a general principle in how TEs are oriented relative to chromatin structures. Future investigations could focus on understanding the mechanisms behind the observed strand bias, the functional implications of the subtle LP-proximal variations, and how the distribution patterns of Alu elements complement or contrast with those of other TE families in shaping genome architecture.

## Additional Transposable Elements

Our analysis of transposable element (TE) distribution around chromatin ligation points (LPs) revealed consistent patterns across multiple TE families and subfamilies. We examined the density of six major TE families: Alu, L1, L2, MER, MIR, and MLT elements, as well as several subfamilies within these groups, within a 100 kb window centered on LPs identified through Micro-C experiments (Fig. 1 for L1, Fig. 2 for Alu; see Supplementary Figures S_ALU to S_MLT for all TE families, and S_AluJr to S_MIRc for specific subfamilies).

The distribution patterns observed for L1 elements (Fig. 1) were largely representative of the trends seen across all examined TE families. Key findings include:

1. Strand Bias: All TE families and subfamilies exhibited a consistent strand bias, with higher density on the plus strand compared to the minus strand. This asymmetry suggests a potential relationship between TE orientation and chromatin structure that extends across TE types and subtypes.
2. Distribution Patterns: While the overall trends were similar, there were some family-specific and subfamily-specific nuances in distribution patterns:
    - L1 elements showed pronounced depletion immediately adjacent to the LP, followed by density peaks at varying distances (Fig. 1, S_L1).
    - L2 elements and its subfamilies (L2a, L2b, L2c) showed similar patterns to L1, but with some variations in the degree of depletion and peak locations (Fig. S_L2, S_L2a, S_L2b, S_L2c).
    - Alu elements as a family displayed a more uniform distribution across the 100 kb window, with less dramatic depletion near the LP (Fig. 2, S_ALU). However, analysis of Alu subfamilies (AluJr, AluSg, AluSx) revealed differences in their distribution patterns, suggesting subfamily-specific interactions with chromatin structure (Fig. S_AluJr, AluSg, S_AluSx).
    - MIR elements and its subfamilies (MIRb, MIRc) showed patterns similar to Alu, with relatively uniform distribution and minor fluctuations (Fig. S_MIR, S_MIR subfamily, S_MIRb, S_MIRc).
    - MER and MLT elements showed intermediate patterns, with moderate depletion near the LP and less pronounced peaks compared to L1/L2 (Fig. S_MER, S_MLT).
3. Reproducibility: The general patterns of TE distribution were consistent between Dataset 1 and Dataset 2 for all TE families and subfamilies, albeit with different overall densities.

This reproducibility across datasets and TE types strengthens the biological significance of our observations. At the same time every TE family and subfamily has substantial differences in peak patterns, suggesting sequence-specificity of the patterns.
4. All the patterns for all the TE families and subfamilies were asymmetric relative to the ligation points, suggesting that the patterns are oriented relative the chromosome direction. (All the graphs have the start of the chromosome on the left and end on the right. This orientation aligns with the standard genomic convention where the short arm (p) of the chromosome is positioned towards the left.)

The consistency of these patterns across two datasets and two ligation points for each of the major TE families and subfamilies suggests a fundamental relationship between transposable elements and chromatin structure that transcends the specific characteristics of individual TE types. The observed strand biases and dataset-specific variations appear to be general features of how TEs are distributed relative to chromatin interaction sites.

These findings collectively point to a complex interplay between TEs and higher-order chromatin structure. The similarities across TE families suggest common mechanisms influencing their distribution, while the subtle differences between families and subfamilies hint at potential specific roles or influences on chromatin organization that may have evolved over time.

## Discussion

The patterns of TE distribution around chromatin ligation points revealed by our study provide several intriguing insights into the potential relationship between TEs and chromatin structure. The consistent strand bias observed across TE families suggests a fundamental principle in how TEs are oriented relative to chromosome arms. This bias could be driven by factors such as transcriptional activity, DNA replication timing, or specific protein-DNA interactions, and warrants further investigation into the mechanisms underlying TE insertion and retention [1].

The distinct distribution patterns observed for different TE families, such as Alu and L1 elements, highlight the importance of considering TE families individually when studying their potential roles in genome organization. The more stable distribution of Alu elements might indicate a consistent structural role for these shorter, more numerous elements, while the variable distribution of L1 elements could suggest a more dynamic interaction with chromatin structure, possibly related to their larger size and lower copy number [2].

The subtle depletion of TEs, particularly L1 elements, near ligation points is an intriguing finding that could indicate distinct properties of chromatin interaction sites. Alternatively, it might suggest that the presence of certain TEs influences the likelihood of chromatin interactions forming at specific locations [3]. This observation aligns with emerging views of genome organization as a hierarchical process involving interactions at multiple scales [4] .

The differences observed between datasets underscore the importance of experimental replication in studies of genome organization. Future studies should aim to expand the number of datasets analyzed to better distinguish consistent patterns from experimental noise.

Our findings contribute to the growing body of evidence suggesting that transposable elements are not merely passive components of the genome but may play active roles in shaping genome architecture. The non-random distribution of TEs around chromatin ligation points supports the hypothesis that TEs are intimately involved in chromatin organization, either as drivers of structural features or as elements preferentially retained in certain chromatin contexts.

# Conclusion

This analysis of transposable element distribution around chromatin ligation points reveals distinct patterns that suggest a functional relationship between TEs and higher-order chromatin structure. The observed strand biases, TE family-specific patterns, and differences from random genomic distributions challenge the notion of TEs as genomic parasites and instead point to their potential role as active participants in genome organization.

Our study opens new avenues for research into the role of transposable elements in genome function and evolution. Future work should focus on elucidating the mechanisms underlying the observed distribution patterns and investigating their potential implications for gene regulation and cellular function. Exploring these patterns across different cell types, developmental stages, and organisms could provide valuable insights into the evolutionary conservation and functional significance of TE-chromatin interactions [5].

As our understanding of genome organization continues to evolve, it is becoming increasingly clear that a comprehensive view of genomic function must include consideration of transposable elements. This work lays the foundation for future investigations that may ultimately reshape our understanding of genome biology and the role of transposable elements in cellular function.

## Acknowledgments

We thank Daniel Oliveira (DO) for genomic computation. The work and the DNA Resonance Research Foundation (DRRF) were funded solely by Max Myakishev-Rempel.

## Author contributions

AV, DO and IS were part-time remote contractors at DRRF. AV and DO did genomic computation. All authors contributed to the discussion.

# References

1. Sultana, T.; Zamborlini, A.; Cristofari, G.; Lesage, P. Integration Site Selection by Retroviruses and Transposable Elements in Eukaryotes. *Nat. Rev. Genet.* **2017**, *18*, 292–308, doi:10.1038/nrg.2017.7.
2. Feschotte, C.; Pritham, E.J. DNA Transposons and the Evolution of Eukaryotic Genomes. *Annu. Rev. Genet.* **2007**, *41*, 331–368, doi:10.1146/annurev.genet.40.110405.090448.
3. Chuong, E.B.; Elde, N.C.; Feschotte, C. Regulatory Activities of Transposable Elements: From Conflicts to Benefits. *Nat. Rev. Genet.* **2017**, *18*, 71–86, doi:10.1038/nrg.2016.139.
4. Bouwman, B.A.M.; de Laat, W. Getting the Genome in Shape: The Formation of Loops,

Domains and Compartments. *Genome Biol.* **2015**, *16*, 154, doi:10.1186/s13059-015-0730-1.
5. Bourque, G.; Burns, K.H.; Gehring, M.; Gorbunova, V.; Seluanov, A.; Hammell, M.; Imbeault, M.; Izsvák, Z.; Levin, H.L.; Macfarlan, T.S.; et al. Ten Things You Should Know about Transposable Elements. *Genome Biol.* **2018**, *19*, 199, doi:10.1186/s13059-018-1577-z.

# Supplement

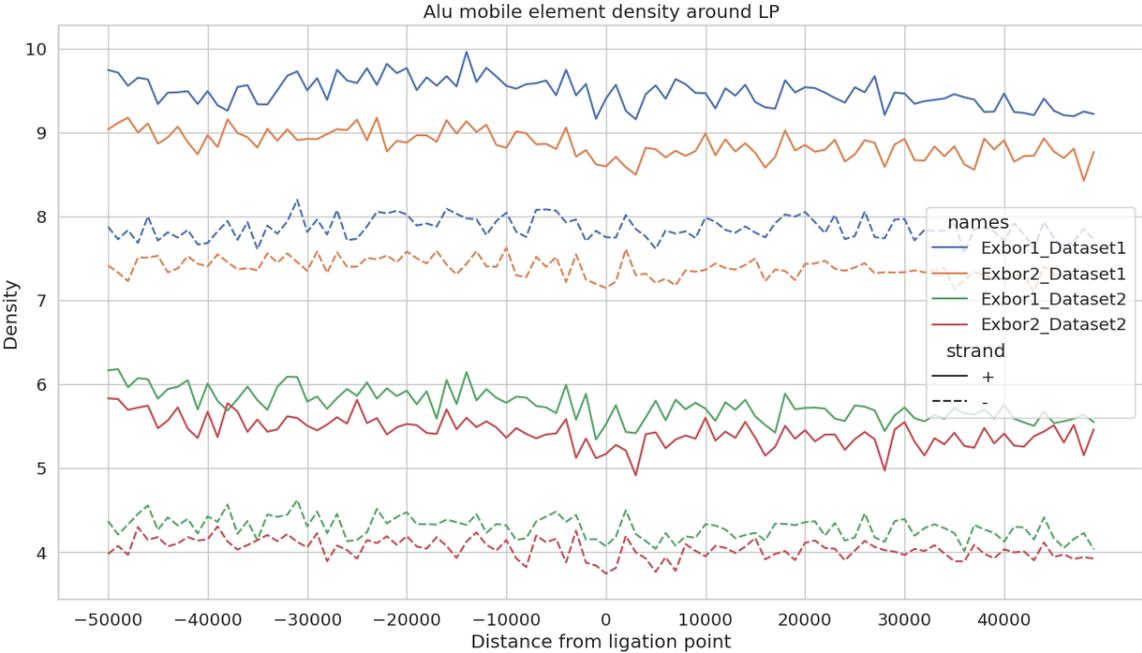

**Fig. S_ALU**

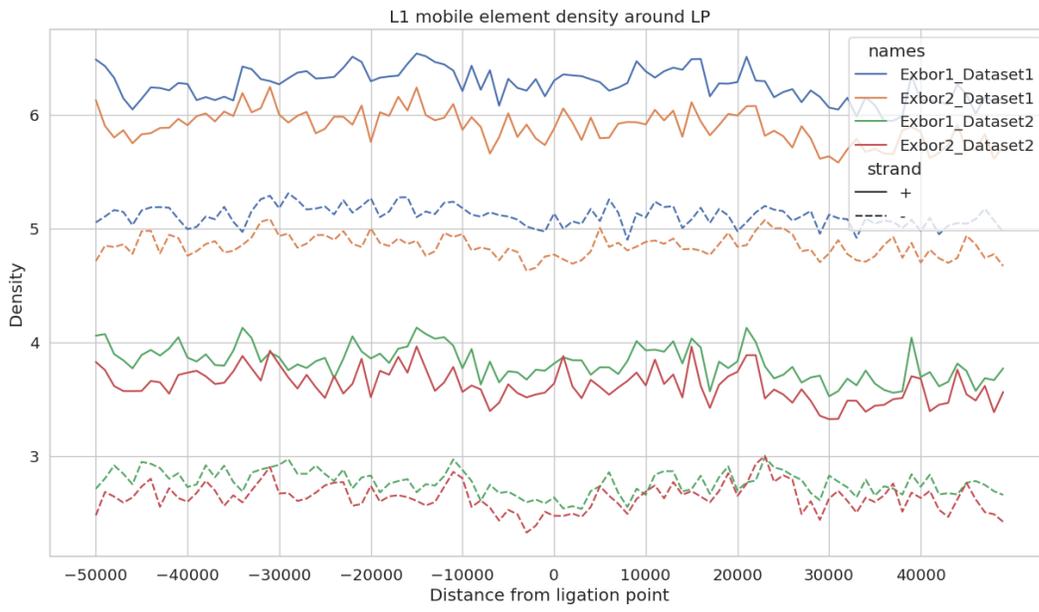

**Fig. S_L1**

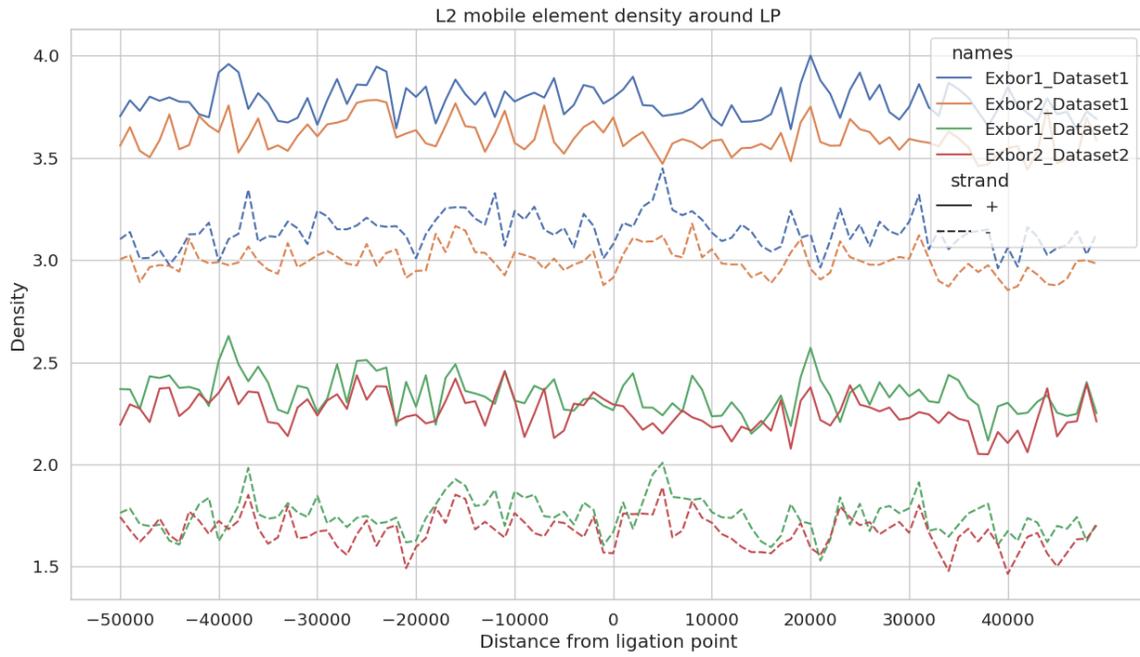

**Fig. S_L2**

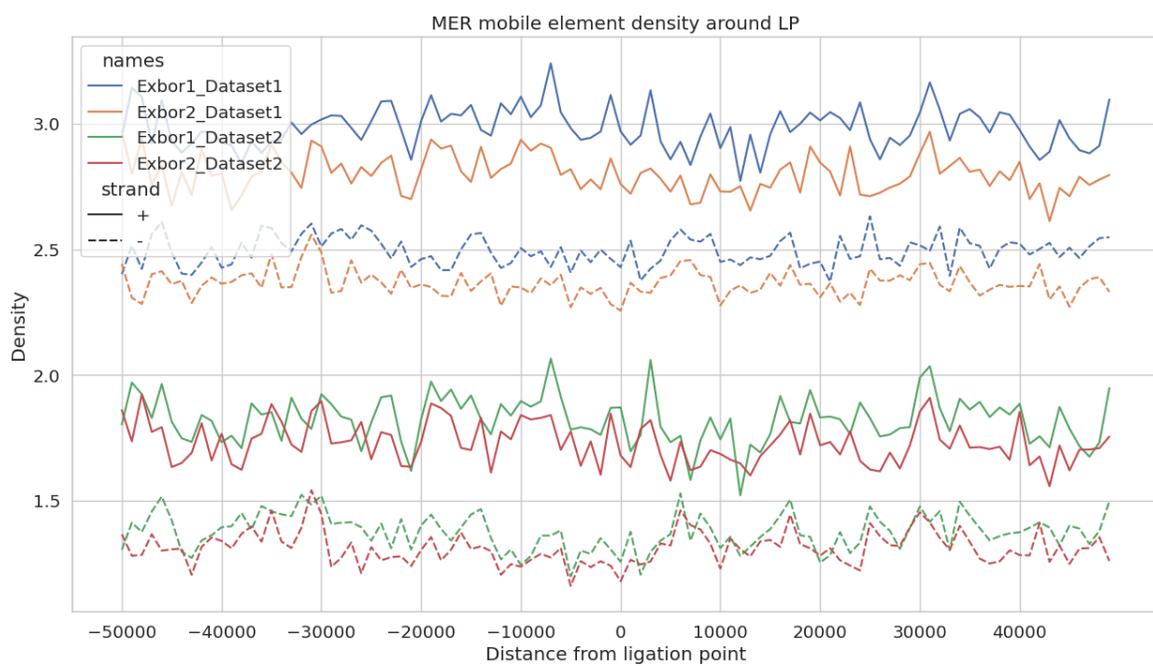

**Fig. S_MER**

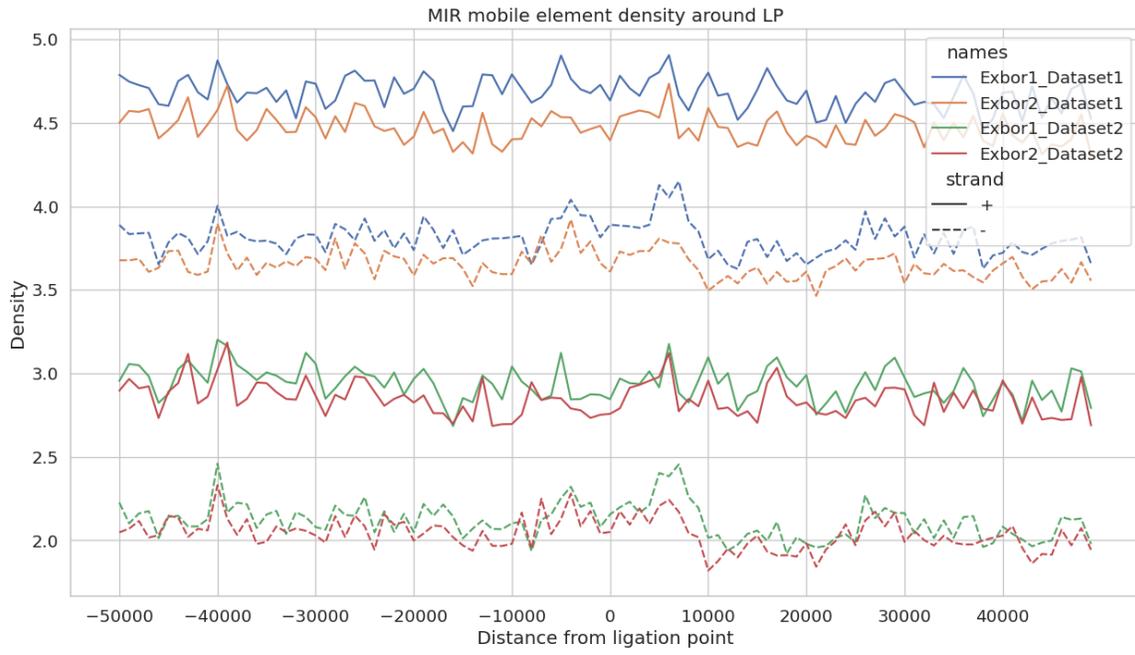

**Fig. S_MIR**

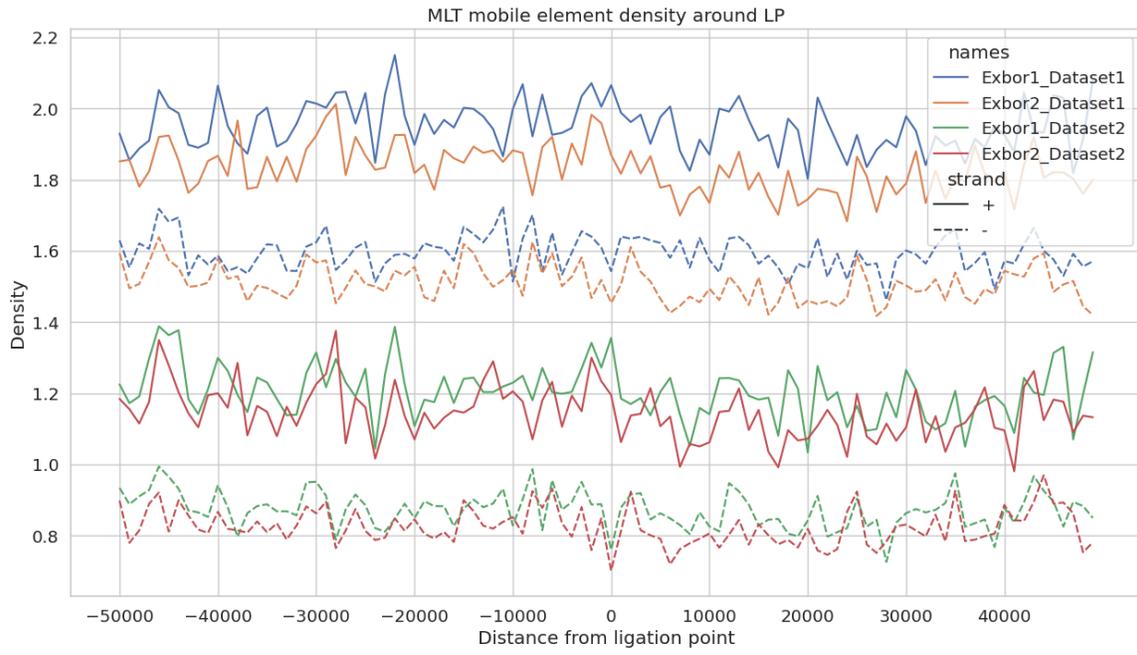

**Fig. S_MLT**

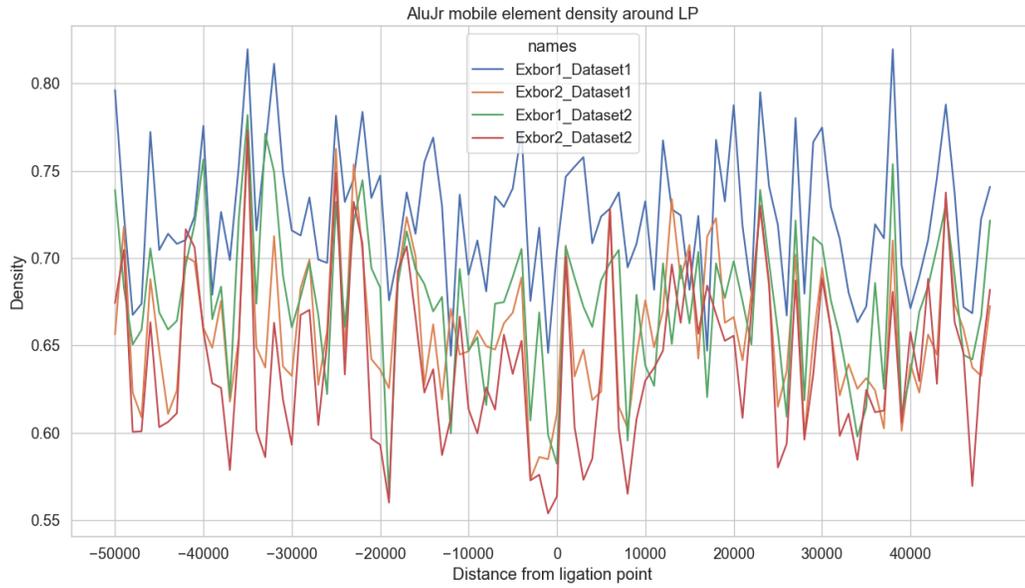

**Fig. S_AluJr**

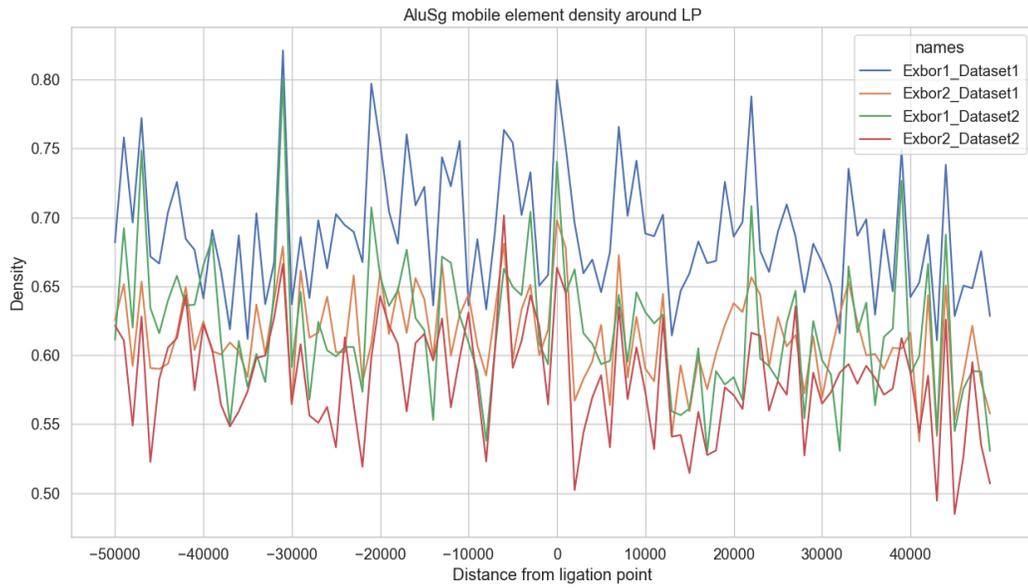

**Fig. AluSg**

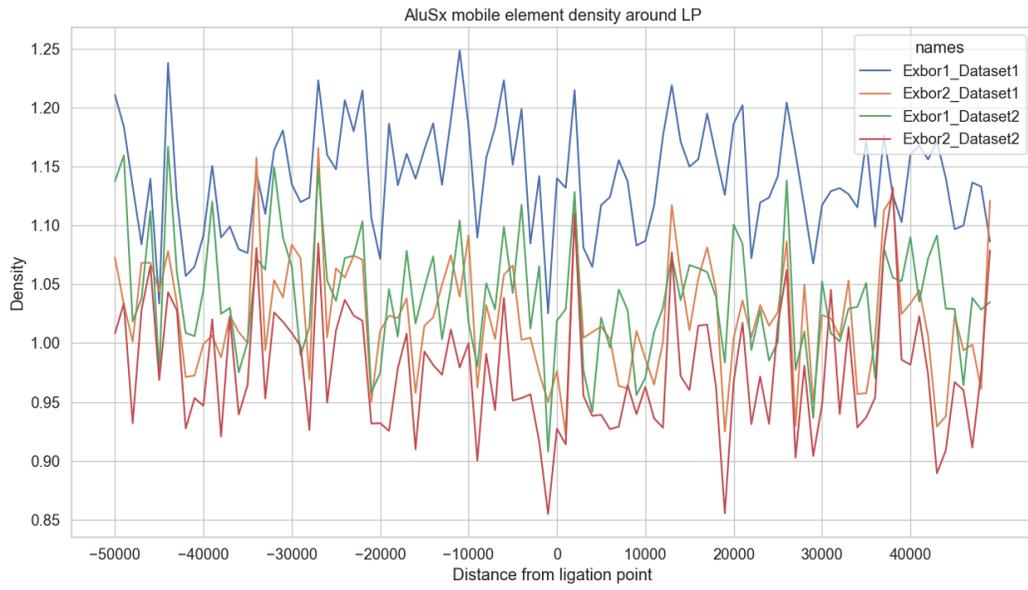

**Fig. S_AluSx**

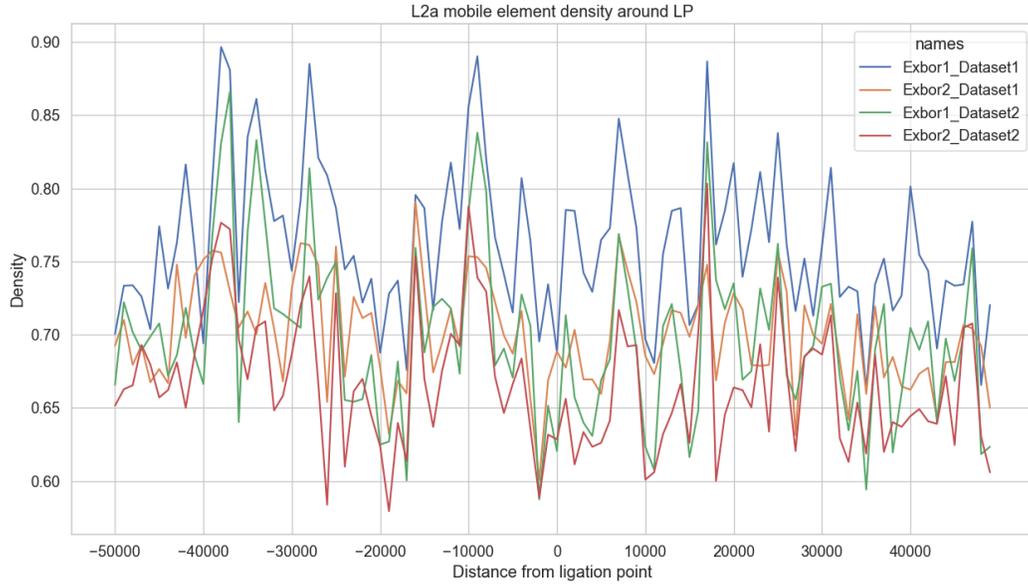

**Fig. S_L2a**

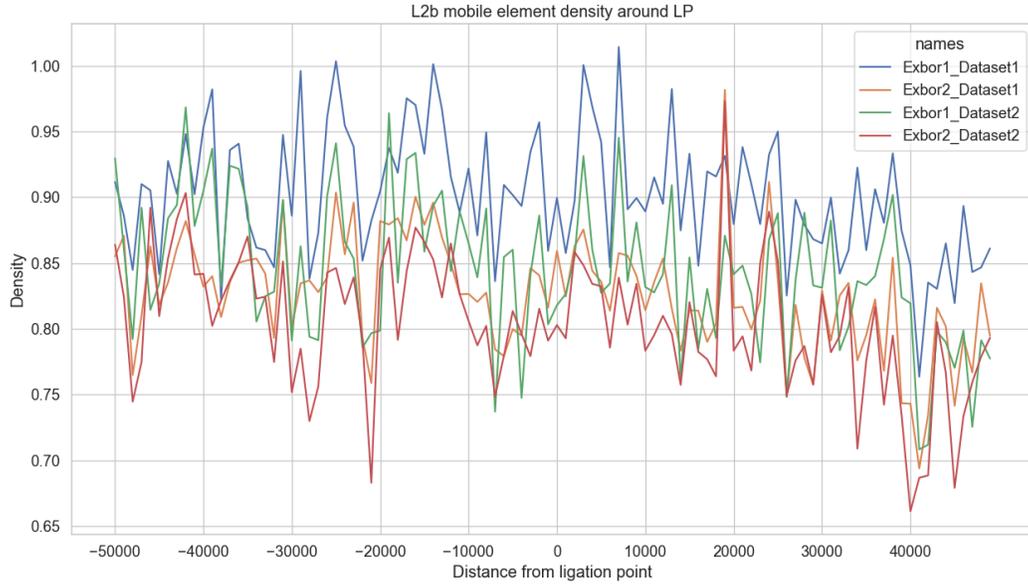

**Fig. S_L2b**

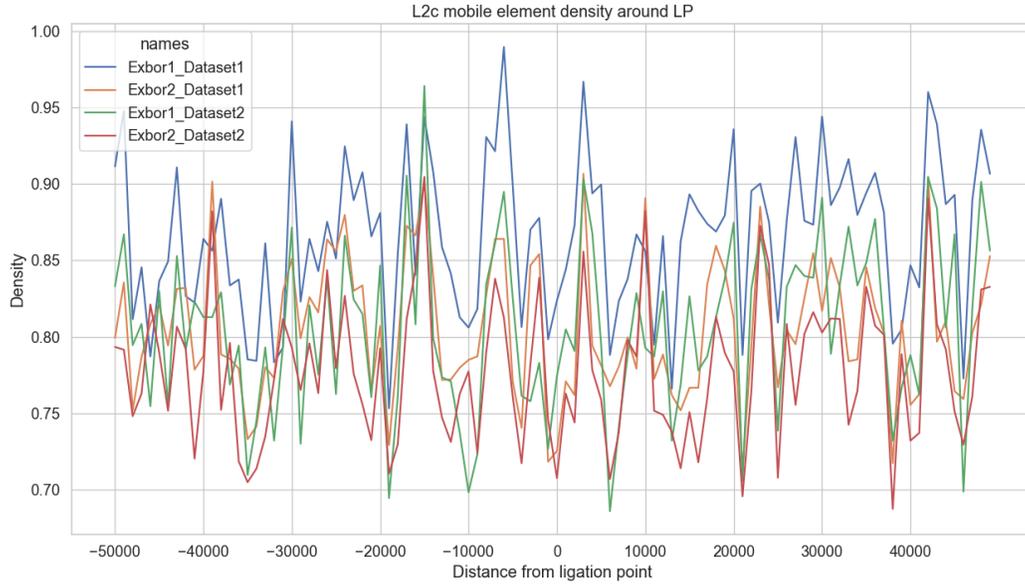

**Fig. S_L2c**

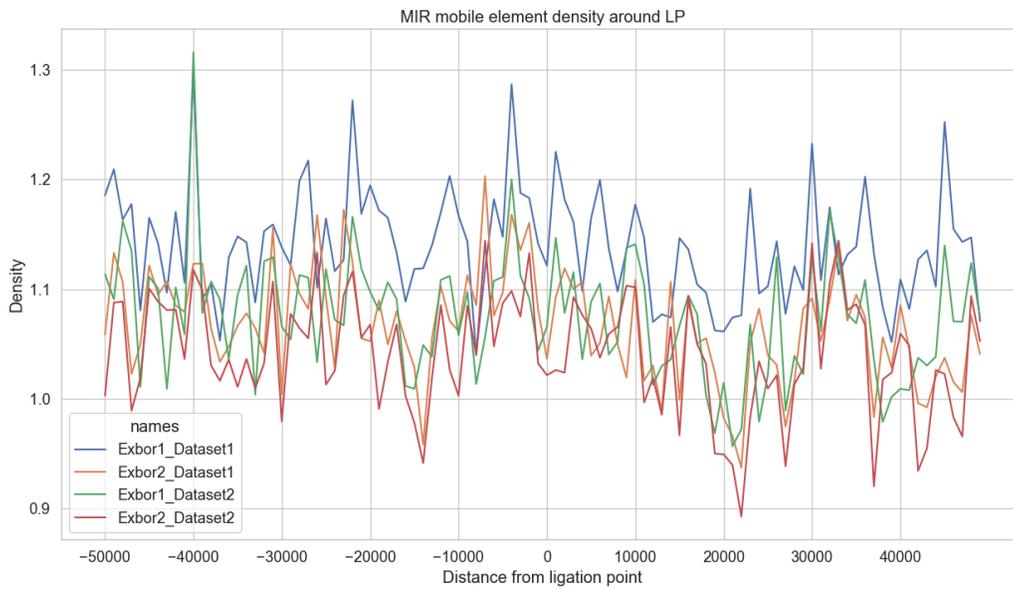

**Fig. S_MIR subfamily**

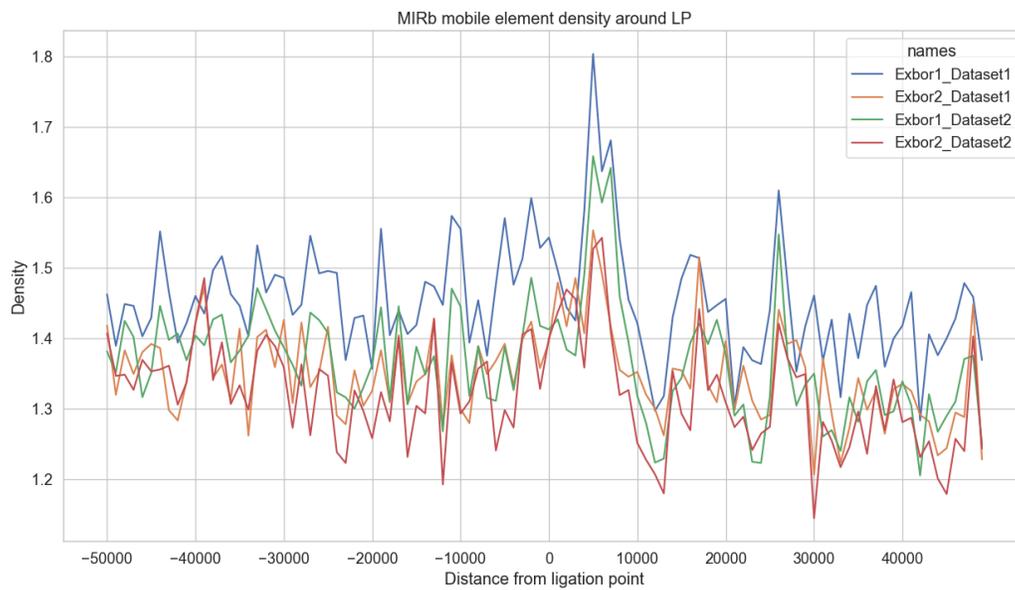

**Fig. S_MIRb**

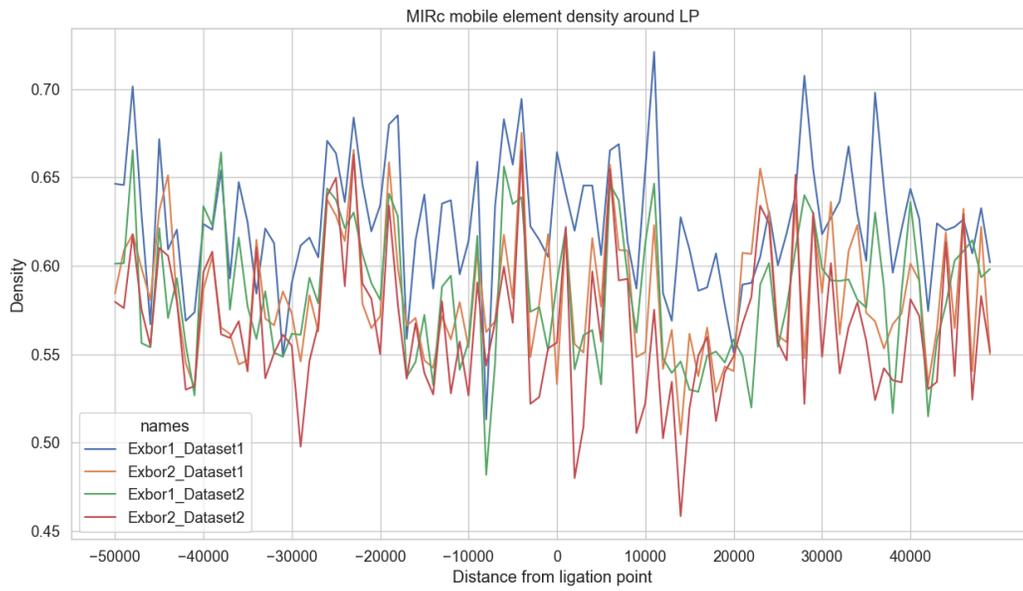

**Fig. S_MIRc**